\documentclass[amsmath,amssymb,aps,prx,twocolumn,groupedaddress,superscriptaddress,10pt,reprint,longbibliography]{revtex4-2}
\usepackage{graphicx}
\usepackage{textcomp} 

\usepackage{exscale}

\usepackage{relsize}

\usepackage[colorlinks,linkcolor=blue,citecolor=blue,urlcolor=blue]{hyperref}

\begin{document}

\preprint{APS/123-QED}

\title{Sunlight-Excited Spontaneous Parametric Down-Conversion for Quantum Imaging}

\author{Ye Xing}
 \altaffiliation[]{These authors contributed equally to this work.}
\author{Deifei Xu}
 \altaffiliation[]{These authors contributed equally to this work.}
\author{Yuan Li}
\author{Wuhong Zhang}
 \email{zhangwh@xmu.edu.cn}
 \author{Lixiang Chen}
 \email{chenlx@xmu.edu.cn}
 \affiliation{Department of physics, Xiamen University, Xiamen, 361005}

\date{\today}

\begin{abstract}
Quantum imaging, which harnesses quantum correlations to achieve imaging with multiple advantages over classical optics, has been in development for several years. Here, we explore sunlight, serving as the pump beam, to excite spontaneous parametric down-conversion to get the quantum correlation of two photons. Remarkably, our investigations disclose that the photon pairs produced from sunlight are well correlated in position such that they can be used for quantum imaging. Consequently, this demonstrates a latent application scenario in which the incoherent beam is harnessed as the pump source for quantum imaging. Our research is of substantial significance as it broadens the scope of available illumination options, such as using scattering light or non-traditional artificial incoherent light sources, for quantum information, a prime potential application being a space-based quantum information mechanism where this approach allows the system to operate independently of a laser.

\end{abstract}

\maketitle


The construction of a global quantum communication network requires the establishment of ultra-long-distance quantum links \cite{vallone2015experimental,bacco2013experimental,liao2017long,yin2020entanglement,deng2019quantum,chen2021integrated}. However, the primary methods of traditional quantum communication—fiber optic links \cite{frohlich2013quantum,sasaki2011field} and free-space optical links \cite{ma2012quantum} are constrained by optical attenuation, Earth's curvature, atmospheric attenuation, and turbulence, making long-distance transmission challenging and hindering the development of a global network. Although the introduction of quantum repeaters can extend communication distances, it significantly increases the complexity of the system. In contrast, satellite-based quantum communication demonstrates immense potential for building a global network and has been experimentally validated in various studies. In recent years, researchers \cite{liao2017long,yin2020entanglement,deng2019quantum,chen2021integrated} have successfully achieved free-space quantum key distribution via satellites, marking a critical breakthrough toward realizing a global quantum communication network. The core component enabling such satellite quantum information is a quantum light source that employs spontaneous parametric down-conversion (SPDC) in nonlinear crystals pumped by lasers to generate entangled photon pairs. Interestingly, some researchers exploited a commercial light-emitting diode (LED), which can be seen as a completely incoherent pump beam, to induce SPDC and examine the properties of the associated biphoton pairs \cite{Tamosauskas2010oe, Galinis2011oe, Galinis2012oc, Nishii2019apl, LiCheng2023pra}. In addition, by using a rotating diffuser to disperse the phase of the laser into a partially spatially coherent beam (also named pseudothermal light), the researchers experimentally achieved the generation of spatial \cite{Hugo2019pra} and polarization \cite{Ismail2017scirep} entangled photon pairs. We also used pseudothermal light and LED as a pump to study the influence of transverse coherence on the two-photon position-momentum, angle-orbital angular momentum correlation \cite{Zhang2019oe,zeng2021controlling}, and further achieved polarization entanglement \cite{Zhang2023prapp} in SPDC with the LED pump. Photon pairs via SPDC from partially coherent pumping have proven to be more robust to the effects arising from atmospheric turbulence \cite{Qiu2012applphysb,Gbur2014joptsocama,Phehlukwayo2020PhysRevA}, and to be used for boosting the two-photon entanglement \cite{hutter2020boosting}. Although the above previous studies have shown that it is capable of obtaining photon pairs via SPDC from the incoherent pump, a truly incoherent source such as sunlight, especially the technique of experimental implementations, has not been studied yet.

Sunlight, as a natural incoherent light source, possesses significant advantages in the space environment due to its abundant availability and ease of access. Currently, its applications have expanded from the traditional field of photosynthesis \cite{he2024structural} to diversified technological domains such as thermal energy conversion \cite{lin2020structured} and photovoltaic power generation \cite{wu2024silicon}. In particular, quantum interference between quantum dot single photons and sunlight was achieved \cite{deng2019quantum}, pushing quantum optic experiments to astronomical scales. Given the pivotal role of quantum light sources in global quantum communication networks, and the demonstrated potential of sunlight in quantum information applications, a critical question naturally arises: can sunlight be experimentally utilized as a pump to generate photon pairs? The challenge of using the photon pairs via SPDC from sunlight pumping is how to make the "movable" sunlight steady into the nonlinear crystal and then realize detectable photon counts with stable fibers connected with the detector. The key advantage of sunlight pumping lies in eliminating the reliance on conventional lasers and external power supplies for space-based quantum information.

\begin{figure*}[t]
\centering
{\includegraphics[width=1\linewidth]{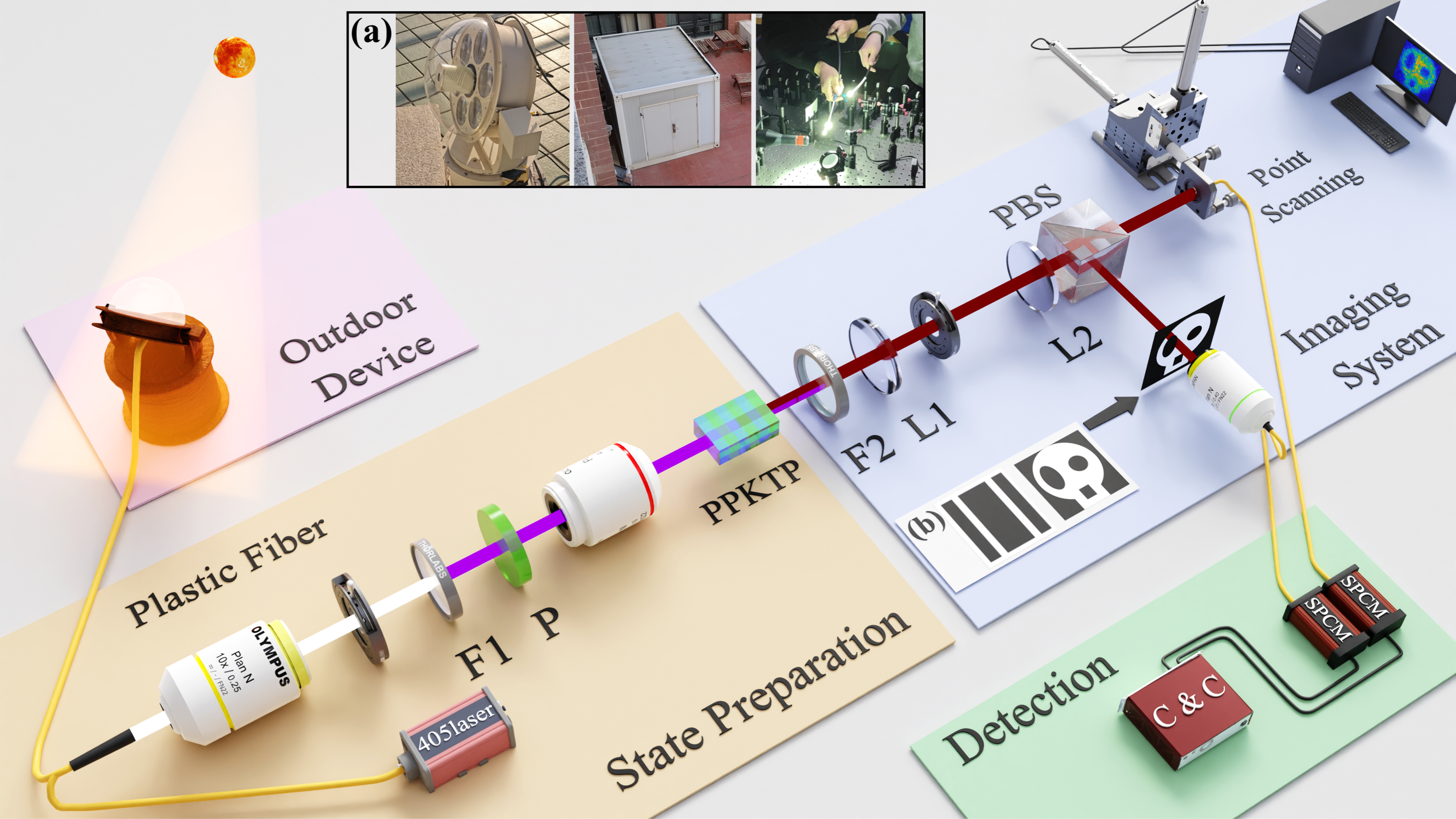}}
\caption{\label{fig:wide}Experimental apparatus for the realization of sunlight-based quantum imaging. Pumping a nonlinear crystal (PPKTP) with sunlight enables the generation of sufficient position-correlated photon pairs, thereby making it possible to implement quantum imaging in a manner similar to that in a laser-based quantum imaging system. (a) Sunlight Collector, outdoor darkroom, the output sunlight from plastic fiber. (b) the double slits and ghost face used for imaging.}
\label{fig1m}
\end{figure*}

In this study, we construct a carefully designed experimental setup to capture sunlight, the most prevalent form of natural light. By using sunlight to pump a nonlinear crystal, we achieve the generation of down-converted photon pairs. Under the quasi-phase-matching conditions of the nonlinear crystal, we discover that the broadband spectrum of sunlight is capable of generating a sufficient number of position-correlated photon pairs. The number of these photon pairs is on par with that generated by a single-frequency laser pump. This property is essential for achieving efficient quantum imaging, as all of these photon pairs contribute to the formation of the object's image. We not only successfully image simple objects, such as double slits, with remarkably high contrast but also manage to image a more intricate two-dimensional object: a ghost face. Our experimental setup, with the advantage of eliminating the need for pre-filtering sunlight, attains a high image contrast of around $95\%$, a value on par with that achieved when using a laser as the pump source. Moreover, our experiment employs a distinct mechanism in which sunlight is used as the pump to generate photon pairs through SPDC, and this approach effectively bridges the gap between classical and quantum ghost imaging. The experimental technique employed in our work represents a significant advancement in the field of quantum imaging and holds great promise for the development of a power-free quantum imaging source for further application.

\section{Theoretical Analysis}
We have conclusively demonstrated that the transverse coherence length of the pump has no bearing on the position correlation of the down-converted photon pairs \cite{Giese_2018, Zhang2019oe}. Under the condition of quasi-phase matching within the crystal, the two-photon cross-spectral density function using a partially coherent single-frequency pump beam can be described by $ W(\boldsymbol{r}_{s}, \boldsymbol{r}_{i}, \boldsymbol{r}_{s}^{\prime}, \boldsymbol{r}_{i}^{\prime})$ (see the Supplemental material for detailed derivation), where $\boldsymbol{r}_s, \boldsymbol{r}_i$ are the position vectors of signal and idler photons.
In the 4f imaging system, the transfer functions from the crystal source plane $\boldsymbol{r}_s, \boldsymbol{r}_i$ to the imaging plane $\boldsymbol{r}_1$, $\boldsymbol{r}_2$ are shown as $h\left(\boldsymbol{r}_s, \boldsymbol{r}_1 \right)= T\left(\boldsymbol{r}_1\right) \delta\left(\boldsymbol{r}_1-M 
 \boldsymbol{r}_s\right)$ and $h\left(\boldsymbol{r}_i, \boldsymbol{r}_2 \right)=  \delta\left(\boldsymbol{r}_2-M\boldsymbol{r}_i\right)$, where $M$ is the imaging magnification. $T\left(\boldsymbol{r}_1\right)$ represents the transmittance function of the object being imaged, corresponding to the double slits and the ghost face in our case. 
 Considering that an objective lens is placed at the imaging object plane $\boldsymbol{r}_1$ to collect photons, the normalized coincidence rate of the two-photon states at another imaging plane $\boldsymbol{r}_2$ can be represented by:
\begin{equation}
\begin{aligned}
R\left(\boldsymbol{r}_2\right) & = \int d \boldsymbol{r}_1 \int d \boldsymbol{r}_s d \boldsymbol{r}_i d \boldsymbol{r}_s^{\prime} d \boldsymbol{r}_i^{\prime} W\left(\boldsymbol{r}_s, \boldsymbol{r}_i, \boldsymbol{r}_s^{\prime}, \boldsymbol{r}_i^{\prime}\right) \\
& \times  h^*\left(\boldsymbol{r}_s^{\prime}, \boldsymbol{r}_1 \right) h^*\left(\boldsymbol{r}_i^{\prime}, \boldsymbol{r}_2 \right) h\left(\boldsymbol{r}_i, \boldsymbol{r}_2 \right) h\left(\boldsymbol{r}_s, \boldsymbol{r}_1 \right) \\
& \propto T\left(\boldsymbol{r}_2\right).
\end{aligned}
\label{eq1}
\end{equation}
From Eq. (\ref{eq1}), we can see that the spatial distribution of the coincidence counts contains information about the imaging object $T(\boldsymbol{r})$, which means that the object can be achieved by scanning the spatial position. The physical interpretation is that for a correlated photon at a specific image-plane position, if it can pass through or reflect off the imaging object, a coincidence count will be detected at the corresponding position. Otherwise, no coincidence counts will occur. It is noted that if a multi-frequency beam is used to pump the nonlinear crystal, it generates non-degenerate broadband down-converted photon pairs \cite{Vanselow2019ol}, and each single-frequency pump is satisfied separately with the Eq. (\ref{eq1}), so that all of these down-converted photon pairs will contribute to the image of the object.

\section{experimental setup and results}
A schematic diagram of the experimental setup that we used in the experiment is shown in Fig. \ref{fig1m}. Firstly, we set up a device outdoors that automatically aligns with the sun throughout the day to collect sunlight. 
The principle of this device is similar to that of an equatorial mount. It automatically adjusts the device's pitch angle based on the current time and geographical coordinates and includes an additional photosensitive chip to enhance alignment accuracy. Once aligned, sunlight is directed vertically onto a 10-cm-diameter Fresnel lens. The light is then coupled into a 20-meter-long plastic multimode optical fiber with a 2.5 mm diameter. To facilitate the experiment, a separate outdoor laboratory (a darkroom) was established specifically for the assembly of the optical setup. 

The other end of the fiber is brought into this laboratory to serve as an experimental light source. The physical picture of our sun collection equipment, outdoor laboratory, and indoor outgoing light spot is shown in Fig. \ref{fig1m}(a). Then a microscope objective (Thorlabs, RMS10X) is used to collect and collimate sunlight, facilitating the setup of the experimental apparatus. A second objective lens (conjugate distance 160 mm, magnification 4X) is followed to focus the collimated beam onto the nonlinear crystal with a numerical aperture (NA) of 0.1. To precisely control the beam size on the crystal surface, an aperture is positioned in front of this objective, reducing the beam diameter to 0.8 mm. The nonlinear crystal is a periodically poled potassium titanyl phosphate crystal (PPKTP), which is phase-matched for type II collinear emission with a size of $1\times 2\times 5 mm^3$. To compare the differences between the sunlight-based quantum imaging system and the laser-based quantum imaging system, we also directly couple a single-frequency laser into the same plastic optical fiber, ensuring that the measurement devices behind the light sources are completely identical. To effectively pump sunlight through the PPKTP, we insert a linear polarizer (P)(LBTEK, FLP25-VIS-M) and a 10 nm spectral filter at 405 nm (F1)(Thorlabs, FBH405-10) into the optical path before the crystal, so that sunlight is completely horizontally polarized with a 10 nm  spectrum, but maintains a small transverse coherence length. A long pass filter (F2) (Thorlabs, FELH0700) is then used to block the pump light. To collect the down-converted photon pairs, we use a 4f system (L1=50 mm, L2=150 mm) to image the crystal plane on the surface of the point detector and the object, as shown in Fig. \ref{fig1m}(b). Another aperture with a diameter of 5 mm is placed in the Fourier plane of the lens L1 to obtain the position correlation of the down-converted photon pairs. The vertically polarized idler photon is reflected by the PBS and passes through the object. Then a microscope objective lens (Thorlabs, RMS20X) collects the photons and couples them into a 0.1 mm diameter multimode fiber connected with the single-photon detector. Similarly, the horizontally polarized signal beam is transmitted through the PBS and is scanned by the point detector. The point detector is composed of a 0.1 mm diameter multimode fiber connected with a single-photon detector. To acquire experimental data, we performed a two-dimensional scan of the multimode fiber in the reflected optical path and set the acceptance window of the coincidence counting circuit as 1 ns. 
Importantly, unlike the typical single-frequency laser pump setup where narrow-bandwidth filters are usually positioned in front of the detectors to ensure that signal and idler photons have the same frequency, our design omits this narrowband filter in front of the detector. This omission allows for broadband position-correlated photon pairs to be utilized for quantum imaging. 

\begin{figure}[t]
\centering
{\includegraphics[width=1\linewidth]{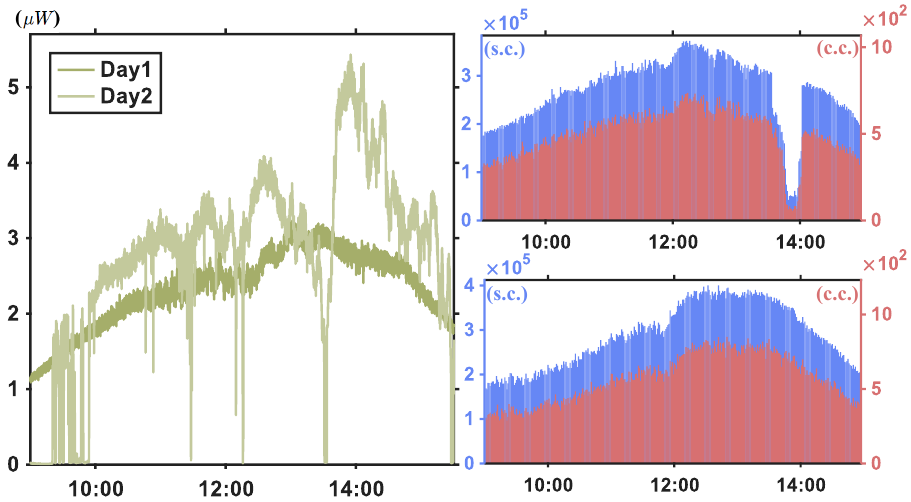}}
\caption{The light intensity distributions and photon count distributions on different separate days. (a) The collected sunlight intensity distribution on two distinct days: the distribution on day 1 corresponds to sunny weather conditions, while day 2 depicts the distribution under cloudy weather. (b) The single (s.c., blue) and coincidence (c.c., red) counts distribution on a day when unexpected cloud obstruction occurred at 14:00. (c) The single (s.c., blue) and coincidence (c.c., red) counts distribution on a clear, cloudless day.}
\label{fig2}
\end{figure}
Firstly, to show the effect of fluctuations in sunlight intensity throughout the day, we measured the pump light intensity using an optical power meter (Thorlabs, S120VC) over approximately 6 hours, with a sampling interval of 1 second. The distribution of the pump light intensity before the crystal in two days from 9 a.m. to 3 p.m. is shown on the left of Fig. \ref{fig2}. One can see that the intensity gradually increases over time, reaching a maximum about 3 $\mu W$ at 1 p.m. for day 1, while day 2 experienced fluctuations at different times. Similarly, to show the distribution of down-converted photons in the same 6-hour period, we measured two different days' single counts and coincidence counts after the SPDC process, as shown in the blue and red curves on the right of Fig. \ref{fig2}. Each measurement was taken every 1 second and accumulated per minute to get enough counts. The coincidence rate for a single count is only about $2.2\text{\textperthousand}$, which is comparable to $3.7 \text{\textperthousand}$ in our previous work \cite{Zhang2019oe} that uses an LED as the pump. It seems that the coherence of sunlight is much worse than that of the LED, which is very challenging for quantum imaging because of the super-short coherence length of the biphotons. So, one needs to carefully position the object as well as the scanning detector in both the signal and the idler arm.

\begin{figure}[t]
\centering{\includegraphics[width=1\linewidth]{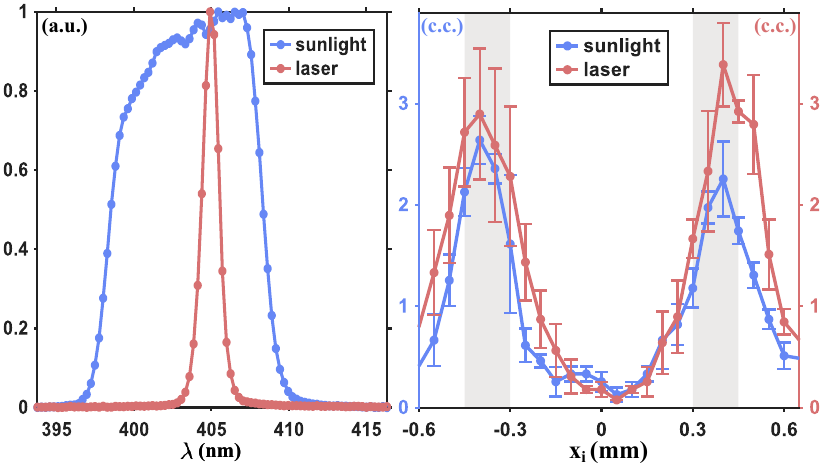}}
\caption{(a) The spectrum of sunlight and the laser after the filter F1. (b) The one-dimensional scanning results of a double slit pumped by sunlight and laser with the approximately same intensity, and the acquisition time for each data point is 13 minutes.}
\label{fig3}
\end{figure}

\begin{figure*}
\centering{\includegraphics[width=1\linewidth]{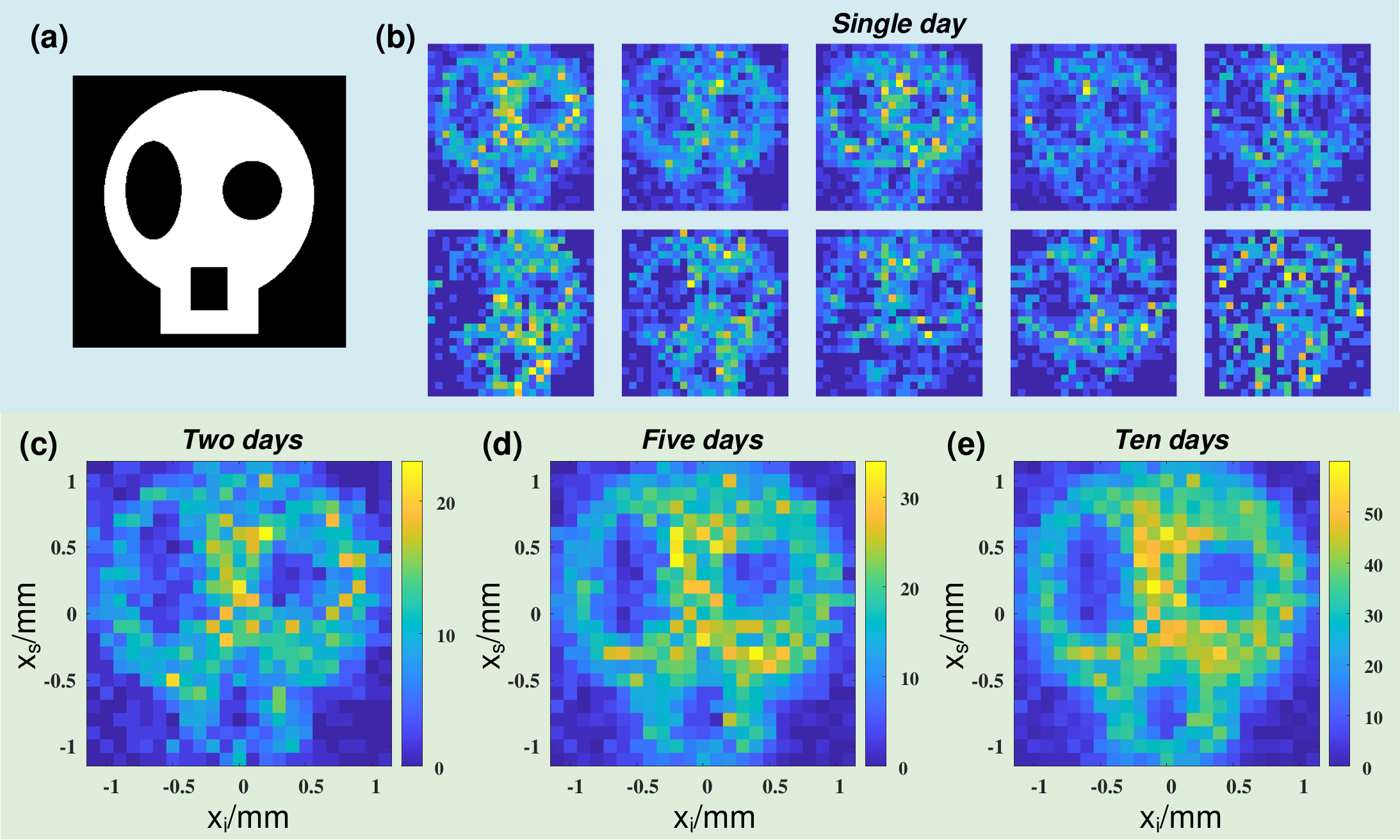}}
\caption{Experimental results of sunlight quantum imaging. (a) Schematic diagram of the ghost face image. (b) Ghost face from a single day, totaling ten days. (c) Ghost face accumulated over 2 days. (d) Ghost face accumulated over 5 days. (e) Ghost face accumulated over 10 days.}
\label{fig4}
\end{figure*}

To demonstrate that the broad-spectrum pump can be used to maintain enough position-correlated biphotons, we compared our sunlight pump with a single-frequency laser pump in a double-slit quantum imaging setup. Fig. \ref{fig3}(a) illustrates the spectrum distribution of both the laser and the sunlight after they pass by a 405 nm filter (F1). The central wavelength of the laser is $\sim$405 nm and the FWHM is $\sim$1.5 nm, while the central wavelength of sunlight is $\sim$403.5 nm and the FWHM is $\sim$10 nm.  Considering that the peak pump light intensity of sunlight is approximately 3 $\mu W$, we also attenuated the laser intensity to about 3 $\mu W$  to ensure consistency in pump light intensity. The double slit, with a slit spacing of 0.75 mm and a slit width of 0.15 mm, is placed in the signal arm. We used a two-dimensional motorized stage to move the multimode fiber head for one-dimensional scanning, with a step size of 0.05 mm and a scanning length of $1.3mm$. Each data point was obtained in 13 minutes for the laser and sunlight pumps, which required approximately 6 hours to complete the scanning.
The averaged results and standard deviations from three sets of experimental measurements are shown in Fig. \ref{fig3}(b), where the blue curve represents the scan result with sunlight, while the red curve represents the scan result with laser. Due to the extremely low pump light intensity, there were only a few coincidence counts per minute. It is well known that the laser pump has much more efficiency than the partially coherent pump, but interestingly enough, we have observed a similar coincidence count level for both cases, indicating that the broadband spectrum of the sunlight pump contributes enough position-correlated photon pairs that are comparable with the single-frequency laser pump. Moreover, one can calculate the contrasts of the double slits, which are about $95\%$ for both cases. This suggests that the position correlation of the biphotons is still effectively preserved, even though the spectrum broadens to some extent. 

Furthermore, to explore the potential of our experimental setup, we replaced the double slit with a more complex two-dimensional image of a ghost face. 
Because the ghost face object, unlike the double-slit structure, cannot reduce the number of scanning points, we must perform a point-by-point scan across the entire two-dimensional plane. 
Additionally, the complex structure of the object minimizes the counts of the bucket detector in the signal arm, requiring a longer integration time to get a clear image.  The image size was approximately $2.3*2.3 mm^2$, with a scanning step size of 0.1 mm, which required $23*23$ points. Each measurement took 1 second, with 40 measurements per point, which led to approximately 6 hours of scanning daily.
Fig. \ref{fig3}(a) displays images from a single day over 10 days. It can be seen that the features of the ghost faces are not particularly prominent. Specifically, some images exhibit relatively distinct features of the ghost faces, which correspond to sunny days with high pump light intensity. In contrast, under less favorable weather conditions, the features of the ghost faces are less pronounced. To get a clearer ghost face image, we can simply add the different days' data.
From Figs. \ref{fig3}(b) to (d), it can be observed that the shape of the image becomes progressively clearer as the number of acquisition days increases. As the photon number gradually increases, the spatial random fluctuations in the photon number gradually diminish, ultimately resulting in a statistically distributed ghost face. After combining the data collected over ten days, the final ghost face imaging result is obtained, as shown in Fig. \ref{fig3}(d). One can see a clear ghost-face structure even by summing different days' data, showing the robustness of our experimental setup. 
Besides, because of the approximately Gaussian distribution of the two-photon state induced by the pump, we can observe some highest counts distributed in the center of the image. The obtained clearly complex ghost face shows the universality and stability of our experimental setup and can be seen as the first quantum imaging with sunlight.

\section{Discussion and conclusion}
 Although it took one day to obtain a ghost image in our experiment, which may seem time-consuming and impractical, it is worth noting that optimizing the sunlight collection system, such as using larger collection lenses and shorter, lower-loss optical fibers, may increase the pump light intensity to the milliwatt level, generating more down-converted photons and enabling faster quantum imaging. Furthermore, research on broadband nonlinear crystals \cite{Fraine2012} or waveguides \cite{Zhang_2024} would help improve the efficiency of down-converted photon generation, which would be beneficial for broadband quantum imaging. On the other hand, employing image reconstruction techniques such as compressed sensing \cite{nc2015} and deep learning \cite{sr2017} could facilitate quantum image recognition at low photon count levels, further reducing the time of the sunlight quantum imaging technique. Our proof-of-principle technology represents a pivotal milestone in sunlight-based quantum imaging. Not only does it provide a novel research platform for this field, it also has the potential to inspire innovative sunlight quantum information applications.

In conclusion, we have built a convenient setup to collect sunlight onto the laboratory to perform the quantum information exploration, which not only realized the generation of correlated photon pairs with sunlight for the first time but also further performed the quantum imaging experiment. Our results show that the desired setup can image not only simple objects, but also complex objects with high visibility. We believe that our technique gives a state-of-the-art to challenge the photon's manipulation and opens the door to explore the sunlight for quantum information. Since the sun is a universally available and free source of natural illumination, our experiments hold promise in demonstrating the feasibility of sunlight as a light source to generate entangled photons \cite{Zhang2023prapp}, which has a wide range of applications, calling for future applications in the establishment of sunlight-pumped quantum light sources in outer space \cite{PhysRevLett.123.080401}.

\begin{acknowledgments}
We appreciate the initial discussions with Professor Robert Boyd, Enno Giese, and Robert Fickler at the early stage of this project. We also appreciate the valuable suggestions from Robert Boyd and Enno Giese for polishing the manuscript. 
This work is supported by the National Key R\&D Program of China (2023YFA1407200 and 2023YFA1407203), the National Natural Science Foundation of China (12034016, 12374280, 61975169), the Natural Science Foundation of Fujian Province (2021J02002, 2023J01007), the Fundamental Research Funds for the Central Universities at Xiamen University (20720220030), and the Xiaomi Young Talent Program.
\end{acknowledgments}

\bibliography{myref}

\end{document}